\newcount\mgnf\newcount\tipi\newcount\tipoformule
\newcount\aux\newcount\piepagina\newcount\xdata
\mgnf=0
\aux=0           
\tipoformule=1   
\piepagina=2     
\xdata=1         
\def\Di{9\ settembre\ 1997}

\ifnum\mgnf=1 \aux=0 \tipoformule =1 \piepagina=1 \xdata=1\fi
\newcount\bibl
\ifnum\mgnf=0\bibl=0\else\bibl=1\fi
\bibl=0

%
%
%
%
\ifnum\bibl=0
\def\ref#1#2#3{[#1#2]\write8{#1@#2}}
\def\rif#1#2#3#4{\item{[#1#2]} #3}
\fi

\ifnum\bibl=1
\openout8=ref.b
\def\ref#1#2#3{[#3]\write8{#1@#2}}
\def\rif#1#2#3#4{}

\fi

\def\9#1{\ifnum\aux=1#1\else\relax\fi}
\ifnum\piepagina=0 \footline={\rlap{\hbox{\copy200}\
$\st[\number\pageno]$}\hss\tenrm \foglio\hss}\fi \ifnum\piepagina=1
\footline={\rlap{\hbox{\copy200}} \hss\tenrm \folio\hss}\fi
\ifnum\piepagina=2\footline{\hss\tenrm\folio\hss}\fi

\ifnum\mgnf=0 \magnification=\magstep0
\hsize=13.5truecm\vsize=22.5truecm \parindent=4.pt\fi
\ifnum\mgnf=1 \magnification=\magstep1
\hsize=16.0truecm\vsize=22.5truecm\baselineskip14pt\vglue5.0truecm
\overfullrule=0pt \parindent=4.pt\fi

\let\a=\alpha\let\b=\beta \let\g=\gamma \let\d=\delta
\let\e=\varepsilon \let\z=\zeta \let\h=\eta
\let\th=\vartheta\let\k=\kappa \let\l=\lambda \let\m=\mu \let\n=\nu
\let\x=\xi \let\p=\pi \let\r=\rho \let\s=\sigma \let\t=\tau
 \let\f=\varphi\let\ch=\chi \let\ps=\psi \let\o=\omega
  \let\D=\Delta

{\count255=\time\divide\count255 by 60 \xdef\oramin{\number\count255}
\multiply\count255 by-60\advance\count255 by\time
\xdef\oramin{\oramin:\ifnum\count255<10 0\fi\the\count255}}
\def\ora{\oramin }

\ifnum\xdata=0
\def\data{\number\day/\ifcase\month\or gennaio \or
febbraio \or marzo \or aprile \or maggio \or giugno \or luglio \or
agosto \or settembre \or ottobre \or novembre \or dicembre
\fi/\number\year;\ \ora}
\else
\def\data{\Di}
\fi

\setbox200\hbox{$\scriptscriptstyle \data $}
\newcount\pgn \pgn=1
\def\foglio{\number\numsec:\number\pgn
\global\advance\pgn by 1} \def\foglioa{A\number\numsec:\number\pgn
\global\advance\pgn by 1}
\global\newcount\numsec\global\newcount\numfor \global\newcount\numfig
\gdef\profonditastruttura{\dp\strutbox}
\def\senondefinito#1{\expandafter\ifx\csname#1\endcsname\relax}
\def\SIA #1,#2,#3 {\senondefinito{#1#2} \expandafter\xdef\csname
#1#2\endcsname{#3} \else \write16{???? ma #1,#2 e' gia' stato definito
!!!!} \fi} \def\etichetta(#1){(\veroparagrafo.\veraformula) \SIA
e,#1,(\veroparagrafo.\veraformula) \global\advance\numfor by 1
\9{\write15{\string\FU (#1){\equ(#1)}}} \9{ \write16{ EQ \equ(#1) == #1
}}} \def \FU(#1)#2{\SIA fu,#1,#2 }
\def\etichettaa(#1){(A\veroparagrafo.\veraformula) \SIA
e,#1,(A\veroparagrafo.\veraformula) \global\advance\numfor by 1
\9{\write15{\string\FU (#1){\equ(#1)}}} \9{ \write16{ EQ \equ(#1) == #1
}}} \def\getichetta(#1){Fig.  \verafigura \SIA e,#1,{\verafigura}
\global\advance\numfig by 1 \9{\write15{\string\FU (#1){\equ(#1)}}} \9{
\write16{ Fig.  \equ(#1) ha simbolo #1 }}} \newdimen\gwidth \def\BOZZA{
\def\alato(##1){ {\vtop to \profonditastruttura{\baselineskip
\profonditastruttura\vss
\rlap{\kern-\hsize\kern-1.2truecm{$\scriptstyle##1$}}}}}
\def\galato(##1){ \gwidth=\hsize \divide\gwidth by 2 {\vtop to
\profonditastruttura{\baselineskip \profonditastruttura\vss
\rlap{\kern-\gwidth\kern-1.2truecm{$\scriptstyle##1$}}}}} }
\def\alato(#1){} \def\galato(#1){}
\def\veroparagrafo{\number\numsec}\def\veraformula{\number\numfor}
\def\verafigura{\number\numfig}
\def\geq(#1){\getichetta(#1)\galato(#1)}
\def\Eq(#1){\eqno{\etichetta(#1)\alato(#1)}}
\def\eq(#1){\etichetta(#1)\alato(#1)}
\def\Eqa(#1){\eqno{\etichettaa(#1)\alato(#1)}}
\def\eqa(#1){\etichettaa(#1)\alato(#1)}
\def\eqv(#1){\senondefinito{fu#1}$\clubsuit$#1\write16{No translation
for #1} \else\csname fu#1\endcsname\fi}
\def\equ(#1){\senondefinito{e#1}\eqv(#1)\else\csname e#1\endcsname\fi}
\openin13=#1.aux \ifeof13 \relax \else \input #1.aux \closein13\fi
\openin14=\jobname.aux \ifeof14 \relax \else \input \jobname.aux
\closein14 \fi \9{\openout15=\jobname.aux} \newskip\ttglue


\font\titolo=cmbx10 scaled \magstep1
\font\ottorm=cmr8\font\ottoi=cmmi7\font\ottosy=cmsy7
\font\ottobf=cmbx7\font\ottott=cmtt8\font\ottosl=cmsl8\font\ottoit=cmti7
\font\sixrm=cmr6\font\sixbf=cmbx7\font\sixi=cmmi7\font\sixsy=cmsy7

\font\fiverm=cmr5\font\fivesy=cmsy5\font\fivei=cmmi5\font\fivebf=cmbx5
\def\ottopunti{\def\rm{\fam0\ottorm}\textfont0=\ottorm%
\scriptfont0=\sixrm\scriptscriptfont0=\fiverm\textfont1=\ottoi%
\scriptfont1=\sixi\scriptscriptfont1=\fivei\textfont2=\ottosy%
\scriptfont2=\sixsy\scriptscriptfont2=\fivesy\textfont3=\tenex%
\scriptfont3=\tenex\scriptscriptfont3=\tenex\textfont\itfam=\ottoit%
\def\it{\fam\itfam\ottoit}\textfont\slfam=\ottosl%
\def\sl{\fam\slfam\ottosl}\textfont\ttfam=\ottott%
\def\tt{\fam\ttfam\ottott}\textfont\bffam=\ottobf%
\scriptfont\bffam=\sixbf\scriptscriptfont\bffam=\fivebf%
\def\bf{\fam\bffam\ottobf}\tt\ttglue=.5em plus.25em minus.15em%
\setbox\strutbox=\hbox{\vrule height7pt depth2pt width0pt}%
\normalbaselineskip=9pt\let\sc=\sixrm\normalbaselines\rm}
\catcode`@=11
\def\footnote#1{\edef\@sf{\spacefactor\the\spacefactor}#1\@sf
\insert\footins\bgroup\ottopunti\interlinepenalty100\let\par=\endgraf
\leftskip=0pt \rightskip=0pt \splittopskip=10pt plus 1pt minus 1pt
\floatingpenalty=20000
\smallskip\item{#1}\bgroup\strut\aftergroup\@foot\let\next}
\skip\footins=12pt plus 2pt minus 4pt\dimen\footins=30pc\catcode`@=12
\newdimen\xshift \newdimen\xwidth \newdimen\yshift
\def\ins#1#2#3{\vbox to0pt{\kern-#2 \hbox{\kern#1
#3}\vss}\nointerlineskip} \def\eqfig#1#2#3#4#5{ \par\xwidth=#1
\xshift=\hsize \advance\xshift by-\xwidth \divide\xshift by 2
\yshift=#2 \divide\yshift by 2 \line{\hglue\xshift \vbox to #2{\vfil #3
\includegraphics{#4.ps} }\hfill\raise\yshift\hbox{#5}}} \def\8{\write13}

\def\V#1{{\,\underline#1\,}}
\def\T#1{#1\kern-4pt\lower9pt\hbox{$\widetilde{}$}\kern4pt{}}
\let\dpr=\partial \let\io=\infty\let\ig=\int
\def\fra#1#2{{#1\over#2}}\let\0=\noindent
\def\guida{\leaders\hbox to 1em{\hss.\hss}\hfill}
\def\tende#1{\vtop{\ialign{##\crcr\rightarrowfill\crcr
\noalign{\kern-1pt\nointerlineskip} \hglue3.pt${\scriptstyle
#1}$\hglue3.pt\crcr}}} \def\otto{\
{\kern-1.truept\leftarrow\kern-5.truept\to\kern-1.truept}\ }

\def\pagina{\vfill\eject}\let\ciao=\bye
\def\txt{\textstyle}
\def\st{\scriptscriptstyle}
\def\*{\vskip0.3truecm}

\def\lis#1{{\overline #1}}\def\eg{\hbox{\it e.g.\ }}
\def\ap{\hbox{\it a priori\ }}
\def\ie{\hbox{\it i.e.\ }}

\def\fiat{{}}
\def\\{\hfill\break} \def\={{ \; \equiv \; }}
 \def\atan{{\,\rm arctg\,}}

\ifnum\aux=1\BOZZA\else\relax\fi
\ifnum\tipoformule=1\let\Eq=\eqno\def\eq{}\let\Eqa=\eqno\def\eqa{}
\def\equ{{}}\fi
\def\defi{\,{\buildrel def \over =}\,}

\def\pallino{{\0$\bullet$}}\def\1{\ifnum\mgnf=0\pagina\else\relax\fi}
\def\W#1{#1_{\kern-3pt\lower6.6truept\hbox to 1.1truemm
{$\widetilde{}$\hfill}}\kern2pt\,}

\def\FINE{
\*
\0{\it Internet:
Author's preprints downloadable (latest version) at:

\centerline{\tt http://chimera.roma1.infn.it}
\centerline{\tt http://www.math.rutgers.edu/$\sim$giovanni}

\0Mathematical Physics Preprints (mirror) pages.\\
\sl e-mail: giovanni@ipparco.roma1.infn.it
}}

\def\NN{{\cal N}}

\def\TT{{\cal T}}
\def\NN{{\cal N}}

\def\HH{{\cal H}}

\def\KJ{{\bf K}}\def\dn{{\,{\rm dn}\,}}\def\sn{{\,{\rm sn}\,}}
\def\cn{{\,{\rm cn}\,}}\def\am{{\,{\rm am}\,}}\def\atan{{\,{\rm arctg}\,}}
\def\aa{{\V \a}}\def\nn{{\V\n}}\def\AA{{\V A}}
\def\pps{{\V \ps}}
\def\oo{{\V \o}}\def\nn{{\V\n}}

\def\ndpr{{\kern1pt\raise 1pt\hbox{$\not$}\kern1pt\dpr\kern1pt}}
\def\Ndpr{{\kern1pt\raise 1pt\hbox{$\not$}\kern0.3pt\dpr\kern1pt}}

\fiat

\centerline{\titolo Fast Arnold's diffusion in isochronous systems}
\*

\centerline{\bf G. Gallavotti}
\*
\centerline{\it Universit\`a di Roma 1, Fisica}
\centerline{\Di}
\*
\*
{\it Abstract: an illustration of a mechanism for Arnold's
diffusion following a nonvariational approach and
finding explicit estimates for the diffusion time.}
\*
\0{\it Keywords: Arnold's diffusion, homoclinic splitting,
KAM}
\*
\0{\bf\S1: Introduction.}
\numsec=1\numfor=1\*

Arnold's diffusion has been established for the simple example
proposed by Arnold, [A], following a nonvariational method, [CG], and
variational methods, [Be], [Br]. The nonvariational method yields
estimates that are terribly big; the variational method instead gives
better estimates, ``{\it fast}'',([Be]) and even very good,
``polynomial'', ones ([Br]).

Here I illustrate the method of [CG] by developing it with the aim of
showing the existence of diffusion, without actually constructing time
scales bounds on the diffusing trajectories. This may lead to a
clarification of a method which maintains its interest in spite of the
better estimates coming from variational methods because it is the
only one which, so far, is robust enough to apply to anisochronous
systems.

If explicit estimates are avoided one gains enormously in simplicity:
this kind of approach was probably the one meant in [A] where the
problem was first posed and solved without bothering to give the
(fairly obvious, see \S5) details. What follows applies also to the
Arnold's case, but I prefer to illustrate it in a case that is even
simpler.

Furthermore I show that if a new idea is added to the method of [CG]
then one can get a ``fast'' (still exponential) estimate for the drift
time.
\*

Here we consider hamiltonians $\HH$ with three degrees of freedom
described by coordinates $I\in R, \V A=(A_1,A_2)\in R^2$ and angles
$\f\in T^1,\aa=(\a_1,\a_2)\in T^2$:

$$\HH=\oo\cdot\V A+\fra{I^2}2+ g^2 (\cos\f -1)+\e f(\f,\aa)\Eq(1.1)$$
where $\oo=(\o_1,\o_2)\in R^2$ is a vector with diophantine constants
$C,\t$, \ie such that for all integer components vectors
$\nn=(\n_1,\n_2)$ it is $|\oo\cdot\nn|^{-1}\le C |\nn|^\t$ if
$\nn\ne\V0$; the {\it perturbation} $f$ is supposed to be a (fixed)
trigonometric polynomial of degree $N$: $f(\f,\aa)=\sum_{0<|\nn|<N,
|n|<N} f_{n,\nn}\cos(n\f+\nn\cdot\aa)$.  The subject being fairly well
understood we do not need to be really very careful about units so
that some coefficients in \equ(1.1) have been set equal to $1$.

One can also use the well known Jacobi's hyperbolic coordinates
$p_0,q_0$ to describe the pendulum $\fra12 I^2+g^2(\cos\f-1)$ near the
unstable point $I=\f=0$, see appendix A1. In the new coordinates,
which we denote $p_0,q_0$, the pendulum hamiltonian becomes
$J(p_0q_0)$ with $J'(x)\defi\fra{d J(x)}{dx}=g+\sum_{n=1}^\io g_n x^n\defi
g(x)$ and the total hamiltonian becomes:

$$\HH=\oo\cdot\V A_0+ J(p_0q_0)+ \e f_0(\aa,p_0,q_0)\Eq(1.2)$$
where $\V A_0\= \V A$, $\aa_0=\aa$ and $f_0(\aa,p_0,q_0)=f(\f,\aa)$.

The function $f_0$, still a trigonometric polynomial in $\V\a$, has
the property: $f_0(\aa,p,q)= f_0(-\aa,q,p)=f_0(-\aa,-p,-q)$ and we
shall call {\it parity} the $4$--elements group of transformations
generated by $P_1:(\aa,p,q)\otto (-\aa,q,p)$ and $P_2:(\aa,p,q)\otto
(-\aa,-p,-q)$. We call $P_0,P_1,P_2,P$ the group elements; we say that
$f_0$ has ``even parity''. If $F(\aa,p,q)= -F(P_j(\aa,q,p))\defi
P_jF(\aa,q,p)$, $j=1,2$, we say that $F$ has odd parity: for instance
$\dpr_\aa f_0$ has ``odd parity''.  The $p$ derivative $F$ or the $q$
derivative $G$ of an even function have the property $F=P_1G=-P_2 F,
G=P_1F=-P_2G$. The Jacobi's map in general transforms functions of
$\f,\aa$ with given parity in $\aa,\f$ in the ordinary sense into
functions with the same parity in the $(\aa,p,q)$ variables.
\*

\0{\bf\S2. Invariant tori and nearby flow.}
\numsec=2\numfor=1\*

We look for a change of coordinates $(\V A_0,\aa_0,p_0,q_0)
\otto (\V A,\pps,p,q)$ which
integrates locally \equ(1.2) near the unstable equilibrium of the
pendulum. More precisely so that in the new coordinates the motion is:

$$\V A=const,\ \pps\to \pps+\oo t,\ p\to p e^{-(1+\g) g t},\
q\to q e^{(1+\g) g t}\Eq(2.1)$$
where $g=g(pq),\g=\g(pq)$ and $x\to\g(x)$ is a suitable function
analytic in $x$ near $x=0$ while $g(x)=J'(x)$ (see \equ(1.2). We shall
attempt to write the change of coordinates:

$$\eqalign{
\V A_0=&\V A + \V H(\pps,p,q),\qquad p_0=p+L(\pps,p,q)\cr
\aa=&\pps,\kern2cm q_0=q+\tilde L(\pps,p,q)\defi q+
L(-\pps,q,p)\cr}\Eq(2.2)$$
where $\V H$ has zero $\pps$ average and even parity.  Setting
$\ndpr\defi q\dpr_q-p\dpr_p$ and imposing that
\equ(2.2) and \equ(2.1) verify the equations of motion one gets
the equations:

$$\eqalign{
\big(g(x)\Ndpr + \oo\cdot\dpr_\pps\big)\,\V H=
&-\e\dpr_\aa f_0(\pps,p_0,q_0) -g(x)\g(x)\Ndpr\V H\cr
\big(g(x)+g(x)\Ndpr + \oo\cdot\dpr_\pps\big)\,L=
&-\e\dpr_{q_0} f_0(\pps,p_0,q_0) -(g(x_0)-g(x))p_0+\cr
&+\g(x)g(x) p-g(x)\g(x)\Ndpr L\cr}\Eq(2.3)$$
where $x=pq, \, x_0=p_0q_0,\, p_0=p+L,\, q_0=q+\tilde L$. In fact the
second equation is independent on the first. Both can be shown to
admit, for $\e$ small enough as we shall always suppose below, a
solution analytic in $\e$ and divisible by $\e$. Note that the unknown
are $\V H, L,\g$. A proof is essentially in the basic paper [Ge]: it
follows the Eliasson's method, [E], as developed in [G2], [GG]. A
``classical'' (\ie by quadratic iterations) proof can be derived from
\S5 in [CG] where the harder anisochronous case is detailed.

\*
\0{\bf\S3. Stable and unstable manifolds.}
\numsec=3\numfor=1\*

{}From \equ(2.2) we can read the following facts:
\*

\0(1) Phase space contains a family of invariant tori $\TT(\AA)$
parameterized by $\AA\in R^2$ and obtained by setting $p=q=0$. The
average position of the tori is precisely $\V A$, because $\V H$ has
zero average: average with respect to $\pps$ or to time. This is a
general property of Thirring's models (like \equ(1.1), see [T] and
[G2]) ``twistless tori''.

\*
\0(2) Given $\AA$ and setting $q=0,\,p\ne0$ one obtains a surface
whose points are parameterized by $\pps,p$ and which is a
local piece of the stable manifold $W^s(\AA)$ of $\TT(\AA)$. The
quantity $-(1+\g(x))g(x)$ is the ``Lyapunov exponent'' of
$W^s(\AA)$. Likewise, setting $p=0,\,q\ne0$ one defines a local piece
of the unstable manifold $W^u(\AA)$ of $\AA$ and $(1+\g(x))g(x)$ is
the corresponding exponent.
\*
\0(3) Given $\AA$ and setting $q\ne0,p\ne0$ one parameterizes the rest
of phase space {\it near} $\TT(\AA)$. In this part of phase space the
motion is in some sense very regular.
\*

\0(4) However the motions, just described locally, are globally more
interesting and chaotic. In fact generically $W^u(\AA),W^s(\AA')$ do
intersect transversally if $\AA,\AA'$ are close enough (depending on
$\e$) and $\TT(\AA),\TT(\AA')$ have the same energy. In such cases
$W^u(\AA)\cap W^s(\AA')$ consists of trajectories, or {\it heteroclinic
intersections}, running asymptotically around $\TT(\AA)$ as $t\to-\io$
and around $\TT(\AA')$ as $t\to+\io$.

The symmetry of the problem implies, see [CG], that if $\AA=\AA'$ and
$\f=\p$ then the point of $W^s(\AA)$ with $\f=\p,\,\aa=\V0$ is {\it
homoclinic}, \ie it is on the trajectory $W^u(\AA)\cap W^s(\AA)$.  The
intersection $W^u(\AA)\cap W^s(\AA')$ exists for all $\AA,\AA'$ close
enough and is generically {\it transversal} in the sense that if we
fix $\f=\p$ (or $\f$ to any other value $\ne 0,2\p$) then any pair of
tangents to $W^u(\AA)$ and $W^s(\AA')$ at common points form an angle
$\ge \m>0$; the bound $\m$ depends on $\e$, of course, and it is
generically proportional to $\e$.

\*
\0(5) Finally a definition: Let $\AA_0,\AA_1,\ldots,\AA_\NN$ be a
sequence such that $|\AA_j-\AA_{j+1}|$ is so small that
$W^s(\AA_j)\cap W^u(\AA_{j+1})$ have a transversal heteroclinic
intersection, in the above sense, with intersection angles $\ge\m$ at
$\f=\p$ . We call such a chain a {\it heteroclinic chain} or {\it
ladder}. One finds in various simple examples $\m=O(\e)=O(\NN^{-1})$,
see [CG]: for a general theory of the splitting see [GGM] and appended
references.

\*
We shall prove the following theorem (``Arnold's diffusion'' or
``drift''):
\*
\0{\bf Theorem 1:\ }{\it Let $\AA_0,\AA_1,\ldots,\AA_\NN$ be a heteroclinic
chain: for any $\d>0$ there are trajectories starting within $\d$ of
$\TT(\AA_0)$ and arriving after a finite time $T$ within $\d$ of
$\TT(\AA_\NN)$.}
\*

This theorem is proved in [CG]; I prove it here againg along the lines
of [CG]: the purpose being of showing the conceptual difference with
respect to the variational approaches, which accounts for the
impressive difference in the time scale of $T$ compared with [Br] or
with the estimate in theorem 2 below (see \equ(6.5)).

\*
\0{\bf\S4. Geometric concepts.}
\numsec=4\numfor=1\*

Let $2\,\k>0$ be smaller than the radius of the disk in the $(p,q)$
plane where the functions in \equ(2.2) are defined. We call $\k$ a
``target parameter''.

To visualize the geometry of the problem involving $2$--dimensional
tori and their $3$--dimensional stable and unstable manifolds, in the
$5$--dimensional energy surface, we shall need the following geometric
objects:

\*
\pallino(a) a point $X_i$, heteroclinic between $\TT(\AA_i)$ and
$\TT(\AA_{i+1})$, which has local coordinates, see \equ(2.2),
$X_i=(\AA_i,\pps_i,0,\k)$.
\*

\pallino(b) the equations, at fixed $q=\k$, of the connected part of
$W^s(\AA_{i+1})$  containing $X_i$, in the local coordinates near
$\TT(\AA_i)$; they will be written as:

$$Y_i(\pps)=(\AA^s_{i+1}(\pps),\pps, p^s_{i+1}(\pps),\k)\Eq(4.1)$$
with $|\pps-\pps_i|<\z$ for some $\z>0$ ($i$--independent): it is
$\AA^s_{i+1}(\pps_i)=\AA_i, \, p^s_{i+1}(\pps_i)=0$ because we require
$Y_i(\pps_i)=X_i$. There are constants $F',F$ such that
$|\AA^s_{i+1}(\pps)-\AA^s_{i+1}(\pps_i)|$ and
$\max_{|\pps-\pps_i|=\,fixed}|p^s_{i+1}(\pps)|$ are bounded, for $\z$
small enough, below by $F'|\pps-\pps_i|$ and above by
$F |\pps-\pps_i|$; the constants $F',F$ have size $O(\m)$.

Note that $W^s(\AA_{i+1})$ also contains a part with local equations
$(\AA_{i+1},\pps,p,0)$ which is {\it not} to be confused with the
previous one described by the function $Y_i(\pps)$. This is more
easily understood by looking at the meaning of the above objects in
the original $(\AA,\aa,I,\f)$ coordinates: in a way the first part of
$W^s(\AA_{i+1})$ is close to $\f=0$ and the second to $\f=2\p$. They
can be close because of the periodicity, but they are conceptually
quite different.

\*
\pallino(c) a point $P_i=Y_i(\tilde\pps_i)$ with $|\tilde\pps_i -\pps_i|=r_i$,
where $\tilde\pps_i,r_i$ will be determined recursively, and a
neighborhood $B_i$:
\kern-5pt
$$B_i=\{|\AA-\AA^s_{i+1}(\pps)|<\r_i,\
|\pps-\tilde\pps_i|<\r_i,
\ |p^s_{i+1}(\pps)-p|<\r_i,\ q=\k\}\Eq(4.2)$$
\nobreak
where $\r_i<r_i$ is another length to be determined recursively. If
$\lis g, 2\lis g$ are a lower and upper bound to $(1+\g(x))g(x)$ for
$|x|<4\k^2$, the point $P_i$ evolves in a time $T_i\simeq \lis
g^{\,-1} \log \k^{-1}$ into a point $X'_i$ near $\TT(\AA'_{i+1})$
which has local coordinates $X'_i=(\AA_{i+1},\pps'_i,\k,0)$.
\*
\pallino(d) The points $\x$ of the set $B_i$ are mapped by the time evolution
to points that, at the beginning at least, come close to
$\TT(\AA_{i+1})$ and in a time $\t(\x)$ acquire local coordinates near
$\TT(\AA_{i+1})$ with $p=\k$ exactly: the time $\t(\x)$ is of the
order of $\lis g^{\,-1}\log\k^{-1}$.

If $S_t$ is the time evolution flow for the system \equ(1.1) we write
$S\x=S_{\t(\x)}\x$ (note that $S$ depends also on $i$). Then $S$ maps
the set $B_i$ into a set $SB_i$ containing:

$$B'_i=\{|\AA-\AA_{i+1}|<\fra1E \r_i,\ |\pps-\pps'_i|<\fra1E \r_i,
\ p=\k,\ |q|<\fra1E \r_i\}\Eq(4.3)$$
because all the points in $B_i$ with $\AA=\AA^s_{i+1}(\pps),\, p=
p^s_{i+1}(\pps), \, q=\k$ evolve to points with $\AA=\AA_{i+1}$,
$p=\k$, $q=0$ and $\pps$ close to $\pps'_i$, by the definitions. Here
$E$ is a bound on the jacobian matrix of $S$ (which, being essentially
a flow over a time $O(\lis g^{\,-1}\log\k^{-1})$, has derivatives bounded
$i$--independently: since we suppose that $\e$ is ``small enough'' we
could take $E=1+b\e$ for some $b>0$ if, as often the case,
$|\AA_i-\AA_{i+1}|<O(\e)$).

\*

\0{\bf\S5. The [CG]-method of proof of the theorem.}
\numsec=5\numfor=1\*

Consider the points $Y^s_{i+1}(\pps)\in W^s(\AA_{i+2})$ with
coordinates $(\AA^s_{i+2}(\pps),\pps,p^s_{i+2}(\pps),\k)$. They will
evolve backwards in time so that $\AA$ stays constant, $\pps$ evolves
quasiperiodically hence ``rigidly'', and $p^s_{i+2}(\pps)$ evolves to
$\k$ while the $q$--coordinate evolves from $\k$ to
$q=p^s_{i+2}(\pps)$ (because $pq$ stays constant, see \equ(2.1)). The
time for this evolution is $T_\pps\simeq \lis g^{\,-1}\log
\k|p^s_{i+2}(\pps)|^{-1}\tende{\pps\to\pps_{i+1}}+\io$.

Therefore there is a sequence $\pps^n\ne \pps_{i+1}$ such that $\pps^n\to
\pps_{i+1}$, $p^s_{i+2}(\pps^n)\to0$ $\AA^s_{i+2}(\pps^n)\to
\AA_{i+1}$ {\it and} $\pps^n-\oo T_{\pps^n}\tende{n\to\io}\pps'_i$, as
a consequence of the diophantine properties of $\oo$. So that there is
$\tilde\pps_{i+1}\defi \pps^n$ with $n$ large enough and a
point $P_{i+1}=(\AA^s_{i+2}(\tilde\pps_{i+1}),\tilde\pps_{i+1},
p^s_{i+2}(\tilde\pps_{i+1}),\k)\in W^s(\AA_{i+2})$ (actually
infinitely many) which evolves, backwards in time, from $P_{i+1}$ to a
point of $B'_i$.

Hence we can define $r_{i+1}=|\tilde\pps_{i+1}-\pps_{i+1}|$ and
$\r_{i+1}$ small enough so that the backward motion of the points in
$B_{i+1}$ enters in due time into $B'_i$.  It follows that the set
$B_i$ evolves in time so that all the points of $B_{i+1}$ are on
trajectories of points of $B_i$. Hence all points of $B_{\NN}$ wil be
reached by points starting in $B_0$.

This completes the proof. All constants can be computed explicitly,
even though this is somewhat long and cumbersome, see [CG].  The
result is an extremely large diffusion time $T$ (namely the value at
$\NN$ of a composition of $\NN$ exponentials! at least this is the
estimate I get after correcting an error in \S8 of [CG]: the error is
minor but leads to substantially worse bounds).

{\it Nevertheless the estimate that comes out of the above scheme seems
essentially optimal.} And then the problem is: ``how is it possible
that by other methods (\eg variational methods of [Be],[Br]) one can
get {\it far better} estimates?

A reason may be that the variational methods are less constructive:
less so than the above. The ``fast drifting'' trajectory exists but
there seems to be no algorithm to determine it, not even the sequence
of its ``close encounters'' with the invariant tori that generates drift:
which is in fact {\it preassigned} in the above method. This certainly can
account for a difference in the estimates. In fact the above
construction is far too rigid: we pretend not only that drift takes
place but also that it takes place via a path that visits closely a
{\it prescribed sequence} of tori in an essentially {\it
predetermined} way. In \S6 a less constructive method is proposed and
used to obtain bounds: which, however, are still far from polynomial.
\*
\pagina
\0{\bf\S6. Fast diffusion: elastic heteroclinic chains.}
\numsec=6\numfor=1\*

The following adds a new idea to the method of [CG], exposed in \S5,
allowing us to improve the superexponential estimate of [CG].  Below
$\e$ will be fixed small enough, and $\lis g$ will be a lower bound to
$g(x)(1+\g(x))$, see \equ(2.1).

Let $y\to \V A(y)$, $y\in[0,1]$, $\AA'(y)\defi \fra{d\V A}{dy}\ne\V0$
be such that the tori $\TT(\AA(y))$ have fixed energy. Then
(evaluating the energy at the homoclinic point $\aa=\V0,\f=\p$) one
sees that $\oo\cdot\AA(y)$ is constant so that the line $y\to \V A(y)$
is parallel to $\oo^\perp =(\o_2,-\o_1)$.
\*

Define $y\to \AA(y)$, $y\in [0,1]$, to be a {\it elastic heteroclinic
chain} with flexibility parameters $\b,\th>0$ and splitting $\m$ if:

(i) for all $|y-y'|<\th\m$ there is a heteroclinic intersection
between the stable and unstable manifolds of $\TT(\AA(y))$ and
$\TT(\AA(y'))$ with splitting angles $\ge\m$ at $\f=\p$.

(ii) the intersection matrix $D\defi \m D_o$ at $\f=\p,\aa=\V0$
verifies:

$$(\V w^\perp\cdot D_o^{-1}\V w^\perp)\defi\,\b\ne0,\qquad \V
w^\perp\defi \fra{\oo^\perp}{|\oo|}\Eq(6.1)$$
where $D, D_o$ are $y$--independent (because of isochrony).

(iii) a heteroclinic intersection at $\f=\p$ between $W^s(\AA(y))$ and
$W^u(\AA(y+\d))$ takes place at $\aa_y(\d)=D_o^{-1}\V w^\perp|\AA'|
\th' +O({\th'}^2)$ for $\d=\m\th',\, |\th'|\le \th$, $|\AA'|=|\AA'(y)|$
and:

$$\fra12 |\AA'|\,\b \,|\tilde\th| < |(\aa_y(\d')-\aa_y(\d''))
\cdot\V w^\perp|< 2 |\AA'|\,\b\,| \tilde\th|\Eq(6.2)$$
for all $\d'=\m \th', \d''=\m\th''$ and $|\th'|,|\th''|<\th$ with
$\tilde\th=\th'-\th''$.
\*

\0{\it Remarks:}

(a) thus every sequence $y_0,y_1,\ldots,y_\NN$ with
$|y_i-y_{i+1}|<\th\m$ is a heteroclinic chain in the sense of \S3, and
the theorem proved in \S5 applies to it. A elastic heteroclinic chain
with parameter $\th$ is also elastic with parameter $\th'<\th$. Hence
it is not restrictive to suppose that $\th$ is as small as needed.

(b) condition \equ(6.1) is a transversality property while
\equ(6.2) is just saying that $\th$ is so small that the involved
first order Taylor's expansions are ``good'' approximations (hence it
is a weak condition and it follows from (ii) provided $\th$ is small
enough). The geometrical meaning of \equ(6.1), \equ(6.2) is that when
$y$ varies by $\d$ (so that $\AA(y)$ varies in $R^2$ {\it
orthogonally} to $\oo$ by $O(\d)$), then the heteroclinic intersection
$\aa_y(\d)$ between $W^s(\AA(y))$ and $W^u(\AA(y+\d))$ is away from
$\V0$ {\it in the direction orthogonal to} $\oo$ by $O(\d\m^{-1})$
provided $\d\m^{-1}=\th'$ is small enough.

The same remains true if one looks at the displacement of the
heteroclinic intersection at any other section located away from the
tori by a fixed distance $\k>0$, if $\e$ is small enough. In fact
consider the intersection matrix $D(t)$ evaluated along the
heteroclinic trajectory at a time $t$ after the passage through
$\f=\p$.  From the equations of motion its evolution is
$D(t)=D-\ig_0^t\dpr_{\aa\f}f(\f(t),\oo \t)
\dpr_\aa \D(\t) \,d\t$ where $\D(t)$ denotes the splitting
in the $\f$--coordinates and $\f(t)$ the heteroclinic evolution of
$\f$: \ie $D(t)=D+O(\e^2)$ (while $D=O(\e)$), see (5.5) in [GGM].

In particular if we look at the heteroclinic intersection point
$\pps_y(\d)$ at $q=\k$, on the same heteroclinic trajectory, and
compare it with the position of the homoclinic point $\pps_y(0)$ of
$\TT(\AA(y))$ at $q=\k$ then we can say that, for some constants
$2b_1,2b_0$ (the factor $2$ is just convenient) it is $|\V
w^\perp\cdot(\pps_y(\d')-\pps_y(\d''))|\in
[2b_1\tilde\th,2b_0\tilde\th]$, with $\tilde\th=(\d'-\d'')\m^{-1}$.
The constants $b_0,b_1$ depend on $\k$, which is however prefixed, and
on $\b$.

(c) examples exist; and generically $\m=O(\e)$.

(d) the above definition is a special case of a natural more general
definition relevant for higher dimensions and for anisochronous
systems.  For instance in the case of anisochronous systems in which a
term $\AA^2/2J$, with $J>0$ constant, is added to \equ(1.1) one has to
require that $y\to\AA(y)$ is a simple rectifiable curve and that,
uniformly in $y\in [0,1]$, \equ(6.1) holds with $D$ replaced by the
intersection matrix $D_y$, and $\oo$ replaced by
$\oo(\AA(y))=\oo+\AA(y) J^{-1}$. In higher dimensions one replaces
\equ(6.1) by requiring that the determinant of the matrix $P_y^\perp
D_y^{-1} P_y^\perp$, with $P_y^\perp$ the projection on the plane
orthogonal to $\oo(\AA(y))$, be non zero uniformly in $y$. But in the
anisochronous cases the condition that for all $y$ there is the torus
$\TT(\AA(y))$, called ``no gap condition'', is strongly restrictive
and quite artificial (although it is verified in the example in [A]).

(e) One can free completely what follows from the parameter $\e$ and
discuss everything in terms of $\m$ only. For simplicity we leave the
general formulation to the reader and stay with $\e$ small enough.
Also for simplicity we shall take $|\AA'|=1$, as this is not
restrictive.
\*

\0{\bf Theorem 2:\it\ Suppose that $y\to \AA(y)$ is elastic in the above
sense, then fixed $a,b$ there exist heteroclinic chains
$\AA_0=\AA(y_0),\AA_1=\AA(y_1),\ldots,\AA_\NN=\AA(y_\NN)$ with
$y_0=a,y_\NN=b$ along which the drift time is $O(\m^{-2})$.}
\*

The estimates proceed by performing the construction of \S5 {\it
without} fixing \ap the heteroclinic chain: we construct it
inductively.

Using the notations of \S4 assume that $y_j$ have been constructed
for $j\le i+1$ together with $\tilde\pps_j,r_j,\r_j, B_j,B'_j$ for
$j\le i$. We must define $y_{i+2}, \tilde\pps_{i+1},r_{i+1},\r_{i+1}$
and, as a consequence, $B_{i+1},B'_{i+1}$.

Let $E$ be as in \S5 and let $E'$ be so large that if $T^0=\lis
g^{\,-1}\log E' {E}^{-1}$ the points $\oo t$, $t\in [0,T^0]$, fill the
torus within $\fra12 b_1\th$ (see (b) above for the definition of $b_1$).
This means that $E'$ is very big: $E'=\exp{O( C\th^{-\t-1})}$ if one
uses the estimate that the time needed to a quasi periodic rotation of
the torus with vector $\oo$, diophantine with constants $C,\t$, to
fill within $\d$ the whole torus $T^2$ is $O(C\d^{-\t-1})$ (for
completeness see appendix A2).

Let $X_{i+1}(y)$ be heteroclinic between $\TT(\AA_{i+1})$ and
$\TT(\AA(y))$ for $y\in[y_{i+1}+\fra12\m\th,y_{i+1}+\m\th]$.  Let the
local coordinates of $X_{i+1}(y)$ be $(\AA_{i+1},\pps_{i+1}(y), 0,\k)$
(see \S4 for the notations). Let, see \equ(4.1):

$$Y_{i+1,y}(\pps)=(\AA^s_{i+2,y}(\pps),\pps,
p^s_{i+2,y}(\pps),\k)\Eq(6.3)$$
be the equation of $W^s(\AA(y))$ in the local coordinates around the
torus $\TT(\AA_{i+1})$ near $X_{i+1}(y)$. We may suppose that
$|\AA^s_{i+2,y}(\pps)-\AA_{i+1}|, |p^s_{i+2,y}(\pps)|<b_3\m
|\pps-\pps_{i+1}(y)|$, for some $b_3$ of $O(1)$ and we may suppose
$b_3>1$, for simplicity.

Suppose $r$ small: a first approximation to $\tilde\pps_{i+1}$ will be
a point $\pps_{i+1,y,r}$ at distance $r$ from $\pps_{i+1}(y)$ such
that:

$$\l\,\defi\, |p^s_{i+2,y}(\pps_{i+1,y,r})|\,=\,
\fra12\max_{|\pps-\pps_{i+1}(y)|=r}\,|p^s_{i+2,y}(\pps)|\Eq(6.4)$$
and $\l\in [b_2 \m r,b_3\m r]$, for some $b_2=O(1)>0$, by the
assumption on the splitting. The constants $b_2,b_3$ are large or
small with the ``target'' parameter $\k$, fixed once and for all, see
beginning of \S4). Let $d=b_2/b_3$. {\it Note that $\pps_{i+1,y,r}$ is
defined non constructively}.

As $r$ varies in the range $d\fra{\r_i}{4b_3 E'}<r<\fra{\r_i}{4 b_3
E}$ the point $\pps_{i+1,y,r}$ varies and $\l$ varies by a factor not
smaller than $E'/E$ by our definition of $d$.  Hence the time $T(r)$
necessary in order that the backward evolution of the point
$Y_{i+1,y}(\pps_{i+1,y,r})=(\AA^s_{i+2,y}(\pps_{i+1,y,r}),\pps_{i+1,y,r},$
$p^s_{i+2,y}(\pps_{i+1,y,r}),\k)$ {\it interchanges} the last two
coordinates will vary by an amount $\ge T^0=O(const\,\th^{-\t-1})$,
see the first lines of this proof and \equ(2.1).

This implies, by continuity, that there will be a value $r(y)$ such
that the backward motion of duration $T(r(y))$ of
$Y_{i+1,y}(\pps_{i+2,y,r(y)})$ has $\pps$--coordinate $\hat
\pps_{i+1,r(y),y}$ close within $\fra12 b_1\th+\fra{\r_i}{4b_3
E}$ to the coordinate $\pps'_i$ of the point $X'_i=(\AA_{i+1},
\pps'_i,\k,0)$ (around which the already known set $B'_i$ is
constructed (see \equ(4.3)); \ie closer than $b_1\th$ (as we suppose
that $\fra{\r_i}{4b_3 E}<\fra12b_1\th$ assuming that $\r_i\le\r_1$ and
$\r_1$ is small enough, which will turn out to be not restrictive). We
can even obtain that $\hat\pps_{i+1,r(y),y}$ is on a chosen side of
the line through $\pps'_i$ parallel to $\oo$.

{\it Note that this is just a continuity statement: hence it is non
constructive}; furthermore other continuity statements will follow.
In this sense the analysis is quite close in spirit to the variational
approaches: nothing is really constructive.

We now vary $y\in[y_{i+1}+\fra12\m\th,y_{i+1}+\m\th]$: the point
$\pps_{i+1}(y)$ varies in the direction orthogonal to $\oo$ by
$b_1\th$ at least (see remark (b) and condition \equ(6.1)).

This means that there is $y^*$, in the considered interval, for which
$\hat \pps_{i+1,r^*,y^*}$, having set $r^*=r(y^*)$, {\it is on the
line parallel to $\oo$ off $\pps'_i$} and within a distance $b_1\th
\m$ of it.

Fixed $y^*,r^*$, let $\pps$ rotate on the circle of radius $r^*$
around $\pps_{i+1}(y^*)$. By the definition of $p^s_{i+2,y^*}(\pps)$
its modulus $\l$ will vary at least by a factor $2$ with respect to
its value $\l^*$ at $\pps_{i+1,y^*,r^*}$, see \equ(6.4), (in fact it
will vary between $2\l^*$ and $0$, extremes included).

Hence by suitably adjusting $\pps$ on the circle we can find a point
$\tilde\pps_{i+1}$ such that the point
$Y_{i+1,y^*}=(\AA^s_{i+2,y^*}(\tilde\pps_{i+1}),\tilde\pps_{i+1},
p^s_{i+2,y^*}(\tilde\pps_{i+1}),\k)$ {\it evolving backwards in time}
exchanges the $p,q$ coordinates in a time $T_{i+1}\simeq T(r^*)$, and
$\pps_{i+1,y^*,r^*}-\oo T_{i+1}\= \pps'_i$ (below we shall worry
about the difference $\pps_{i+1,y^*,r^*}\ne \tilde\pps_{i+1}$).

In fact in order to obtain this we only have to change the time
$T(r^*)$ by an amount $O(b_1\th)$, and this is achieved by varying
$\l$ off $\l^*$ by a factor $e^{O(b_1\th)}\in
[\fra12,2]$, as it is not restrictive to take $b_1\th$ small.

Since $\tilde\pps_{i+1}$ differs from $\pps_{i+1,y^*,r^*}$ by a small
amount $2 r^*\le \fra{\r_i}{2b_3E}$ (at most, by construction), this
means that the point $\tilde\pps_{i+1}-\oo T_{i+1}$ differs from
$\pps'_i$ by at most $\fra{\r_i}{2E}$, recalling that $b_3>1$. Since
$r^*<\fra{\r_i}{4b_3E}$ the other coordinates verify
$|p^s_{i+1,y^*}(\tilde\pps_{i+1})|<b_3 \m\fra{\r_i}{2b_3E}$ and
$|\AA_{i+2,y^*}^s(\tilde\pps_{i+1})-\AA_{i+1}|<b_3 \m\fra{\r_i}{2b_3E}$
(recall that $\AA_{i+2,y^*}^s(\pps_{i+1}(y^*))\=\AA_{i+1}$). Hence the
point $Y_{i+1,y^*}$ evolves, backward, in time $T_{i+1}$ to a point well
inside $B'_i$: so do, in a time which differs suitably from $T_{i+1}$ by
a factor of $O(1)$, all the points close enough to it, say within
$\r_{i+1}=r^*/2$.

Hence if we set $r_{i+1}=r^*\ge d\fra{\r_i}{4b_3E'}$ and
$\r_{i+1}=r_{i+1}\fra{\r_i}{4^2E}\fra{\lis g}{ |\oo|}$, $y_{i+2}=y^*$ and
$\AA_{i+2}=\AA(y_{i+2})$ we see that all points of $B_{i+1}$ are on
the forward evolution of points in $ B'_i$. The time needed for the
passage through $B_{i+1}$ of the points of $B'_i$, which visit it, is
bounded proportionally to $T_{i+1}$. The radius $\r_i$ has to be
chosen so small so that the ratio between the variation of the time
to exchange $p$ and $q$ is small enough (\ie $O(\fra{\r_i}{E})$).

It follows that within a time $T=const\,\sum_{i=0}^\NN T_{i+1}$ the
whole chain will be run by some trajectories. Here $\NN\ge 2(b-a)(\th
\m)^{-1}$ and $T_i\le O(\lis g^{\,-1}\log (E')^i)$, so that drift
between $\AA(a)$ and $\AA(b)$ takes place in a time $O(\lis
g^{\,-1} 2^\NN)$:

$$T\,\le\, const\,C \,\lis g^{\,-1}\, 2^{\m^{-1}\th^{-1}}\Eq(6.5)$$
and, recalling that $\th$ is fixed, if $\m=O(\e)$ this is
$const\,2^{const \e^{-1}}$.

\*
\0{\bf\S7. Concluding remarks. Very fast diffusion?}
\numsec=7\numfor=1\*

For a review on diffusion see [L]: in this paper the possibility of
estimates of size of an inverse power of $\e$ is proposed and
discussed.
\*

(1) The above nonvariational proof gives results not directly
comparable to the best known, [Be], [Br], based on a variational
method and giving (in [Br]) a polynomial drift time of $O(\m^{-2})$.

The papers [Be],[Br], deal with Arnold's example, [A], \ie with a
different case. They use in an essential way the structure of the
model, implying existence of a ``gapless'' system of local coordinates
in which the motion is ``trivial'' (\ie given by \equ(2.1)).  Although
such coordinate system does not appear explicitly in the proofs in
[Br], it nevertheless exists under Arnold's assumptions as shown by
[P].

Therefore this difference between the present paper and [Br] is not so
important: the above proofs apply also to the model in [A] and [Br]
(in fact in absence of gaps also anisochronous systems admit
coordinates $(\AA,\pps,p,q)$ with the properties of the ones in
\S2). Only the constants may be affected (in particular the ones in
\equ(6.5)), although I do not think that \equ(6.5) changes. It is hard
to see how to improve the bounds of \S6, which are already quite non
constructive. Hence the difference between the size of the bounds
remains a puzzle that I do not understand.

(2) It is worth stressing that the above methods apply every time
there is a heteroclinic chain and ``no gaps'' around resonant tori:
therefore they apply to the case in [A] with, in the notations of [A],
$\m=\e^c$ and $c$ large enough.

In the isochronous models they apply, immediately, to a variety of
cases: a non trivial one is the hamiltonian \equ(1.1) with
$\oo=(\h^a,\h^{-1/2})$, $a\ge0$, $\e=\m\h^c$ with $c$ large enough
and, {\it possibly, even a further `` monochromatic, strong and
rapid}'' perturbation $\b f_0(\f,\l)$ like $\b \cos(\l+\f)$ with
$\b=O(1)$. Suppose that we consider only values of $\h$ such that
$|\oo\cdot\nn|> C\h^d|\nn|^{-\t}$ for all $\V0\ne \nn\in Z^2$, see \S2
in [GGM].  Then {\it by using the results of} [GGM] (\S8) we see that
if $\h$ is fixed small enough the homoclinic splitting is analytic in
$\b$ for $|\b|<O(\h^{-1/2})$, while it {\it does not vanish} for $\b$
small (\ie $\b= O(\h^c)$), generically in $f$ (but it is very small,
see [GGM], \S6). Hence it is not $0$ for all $\b< 2$ (say) {\it except
possibly finitely many values of $\b$}. This means that in such {\it
strongly perturbed systems} ($\b=O(1)$) one still has elastic
heteroclinic chains of arbitrary length, see \S8 of [GGM], and
therefore there is diffusion (provable by the methods of
\S5,\S6). Furthermore the $\AA$--independent (because of isochrony,
see [GGM]) homoclinic angles can become large when $\b,\m$ approach
their convergence radii and this gives us the possibility of ``very
fast'' drift on time scales of $\sim O(1)$.  In fact I think that the
homoclinic splitting might be a monotonic function of $\e,\b$ for
interesting classes of perturbations.

A similar analysis can be made for the model in [A] (which is also
without gaps).
\*
(3) An advantage of the technique of \S5 is its flexibility which
makes it immediately applicable, essentially without change, to
anisochronous systems, see [CG].
\*

(4) Constructivity, even partial (see comments in \S5), seems the key
to understanding the huge difference between the results of \S5 and
the variational results, or those of \S6 above: diffusion time bounds
in an inverse power of $\e$ (in [Br] and \S6) versus an
exponential in the more constructive proposal in \S5.  A hint in this
direction is provided by the bound in \S6: by adding a new idea to the
method of \S5, \ie of [CG], one can get a drift time estimate of
$O(\e^{-2})$ instead of the exponential of [CG], and \S5. {\it But the
theory becomes now less constructive}: not even the sequence of close
encounters with invariant tori is determined.
\*

(5) It seems possible that the construction of \S6 might be rendered
constructive without losing the bounds: this is certainly an
interesting problem.
\*

(6) Finally we discussed only drift in phase space: but it is clear
that heteroclinic chains do not need to ``advance'' at each step (\eg
a $A$--coordinate needs not to increase systematically): we can use
heteroclinic chains that advance and back up at our prefixed wish (\eg
randomly). Hence, in this sense, there is no difference between drift
and diffusion.
\*\*

{\bf Acknowledgements:\rm\ I am indebted to P. Lochak stimulating
comments and, in particular, to G. Gentile and V. Mastropietro for
many discussions and help in revising the manuscript. This work is
part of the research program of the European Network on: "Stability
and Universality in Classical Mechanics", \# ERBCHRXCT940460.}
\*
\pagina
\0{\bf Appendix A1. Jacobi's map.}
\numsec=1\numfor=1\*

This appendix is standard: here it is taken from A9 of [CG] with small
changes, to use it for future references.

The theory of jacobian elliptic functions shows how to perform a
complete calculation of the functions, below denoted $R,S$, in terms
of which the canonical Jacobi's coordinates are defined, see [GR]
(9.198),(9.153), (9.146), (9.128), (9.197).  The result, reported
for completeness, is discussed in terms of the pendulum
energy:
$${\dot\f^2\over2}+g^2(1-\cos\f)=E\Eqa(A1.1)$$
where the origin in $\f$ is set at the stable equilibrium, to adhere to
the notations in the theory of elliptic functions.
Setting $u=t(E/2)^{1/2}\=\e^{1/2} gt$, $k^2={2 g^2/E}=\e^{-1}$
where $\e$ is the {\it dimensionless} energy so that $\e=1$ is the
separatrix, let:
$$\KJ(k)=\ig_0^{\p/2} {d\a\over (1-k^2\sin^2\a)^{1/2}}\Eqa(A1.2)$$
We shall use the ``standard'' notations (\ie those in
[GR]) for the jacobian elliptic integrals {\it except} for $x(.)$, which is
usually denoted $q(.)$, but which we would confuse with the
variable $q$ that we want to construct:
$$\eqalign{ k'=&(1-k^2)^{1/2},\qquad g_J=g{\p\over2k\KJ(k')}=\e^{1/2}
g,\qquad \l\={1\over2}{1-k^{1/2}\over1+k^{1/2}}\cr
x(k')=&e^{-\p\KJ(k)/\KJ(k')}= \l+2\l^5+15\l^9+150\l^{13}+1707\l^{17}+
\ldots\cr}\Eqa(A1.3)$$

In terms of the above notations we have, directly from the definitions
(\ie from the equations of motion):
$$
I(t)=\dot \f=-2g\e^{1/2}\dn(u,k),\qquad
\f(t)=2\am(tg\e^{1/2})\Eqa(A1.4)$$
which yield, changing the origin for $\f$ to the
unstable point to conform with our notations (\ie obtaining
$\f(t)=2(\am(tg\e^{1/2})+\p/2)$), for $I(t)=R(p(t),q(t)),
\f(t)=S(p(t),q(t))$:
$$R=-2g\e^{1/2}{\dn(iu,k')\over\cn(iu,k')},\quad
\sin {S\over2}={1\over \cn(iu,k')},\quad\cos {S\over2}=
i{\sn(iu,k')\over\cn(iu,k')}
\Eqa(A1.5)$$

Setting $p=e^{-g_Jt}, q=x(k') e^{g_J t}$, see [GR], and using
$R(p,q)=g_J (-p\dpr_p+q\dpr_q)S(p,q)$ to evaluate $S$ from $R$, the
quoted basic relations between elliptic integrals imply
immediately that the $I(t)=R(p(t),q(t)), \f(t)=S(p(t),q(t))$,
solve the pendulum equations if:
$$\eqalignno{
\txt R(p,q)=&\txt -2g_J
\Bigl[{p\over1+p^2}+{q\over 1+q^2}-\sum_{n=1}^\io(-1)^n{1+x^{2n-1}\over
1-x^{2n-1}}(p^{2n-1}+q^{2n-1})\Bigr]\cr
\txt S(p,q)=&\txt 2\left[\atan p-\atan q-\sum_{n=1}^\io(-1)^n{1+x^{2n-1}\over
1-x^{2n-1}}{(p^{2n-1}-q^{2n-1})\over 2n-1}\right]&\eqa(A1.6)\cr
\txt \sin {S(p,q)\over2}=&\txt \fra{g_J}{g}
\Bigl[
{p\over 1+p^2}-{q\over 1+q^2}-\sum_{n=1}^\io(-1)^n{1-x^{2n-1}\over
1+x^{2n-1}}(p^{2n-1}-q^{2n-1})\Bigr]\cr
\txt \cos {S(p,q)\over2}=&\txt -\fra{g_J}{2g}
\Bigl[
{1-p^2\over 1+p^2}+{1-q^2\over 1+q^2}+2
\sum_{n=1}^\io(-1)^n{1-x^{2n}\over
1+x^{2n}}(p^{2n}+q^{2n})\Bigr]\cr}$$
with $x\=pq$.  Note that $g_J$ depends on $x$, and so do $k',k$: hence
the coefficients of the first and of the last two of \equ(A1.6) are
also functions of $x=pq$.  Furthermore the (dimensionless) energy $\e$
becomes a function of $\x=pq$ defined by inverting the map:
$$\x=x(k')\=x((1-\e^{-1})^{1/2})\Eqa(A1.7)$$
and the point corresponding to $\f=\p$ and to a
dimensionless energy $\e$, has coordinates:
$$p\=1,\quad q\=x(k')\Eqa(A1.8)$$
(a rearrangement of \equ(A1.6) showing convergence for $p=1$ and
$|x|<1$ is exhibited below).

The variables $(p,q)$ defined above are nice and natural: however they
are not canonically conjugated to $(I,\f)$: the jacobian determinant
of the map $(p,q)\otto(I,\f)$ is not $1$. But the jacobian determinant
must be a function $D(x)=\fra{\dpr (p,q)}{\dpr(I,\f)}$ of $x$ alone
(\ie of the product $pq$); then \equ(A1.8) and the equations of motion
imply that $D(x)^{-1}= g_J^{-1}\fra{2g^2d\e(x)}{dx}=4g
\fra{d\e^{1/2}}{dx}$.

Therefore one can modify the variables $p,q$ into new variables
$(p_J,q_J)=(p F(x), q F(x))$ with $F$ such that the jacobian
$\fra{\dpr (p_J,q_J)}{\dpr(I,\f)}=\fra{\dpr (p_J,q_J)}{\dpr(p,q)}
D(x)$, which is $D(x)\cdot\dpr_x(x F^2(x))$, is identically $1$. One
finds: $F(x)=(4g)^{1/2}(\fra{\e^{1/2}-1}x)^{1/2}$.

To invert the map $(p_J,q_J)=(p F(x), q F(x))$ define $x_J\defi
p_Jq_J$ and $G(x_J)\defi F(x)^{-1}$ then: $p=p_J G(x_J)$ and $q=q_J
G(x_J)$, $x=x_J G^2(x_J)$.  The final result is a local canonical map
between Jacobi's coordinates $(p_J,q_J)$ and global $(I,\f)$
coordinates:
$$I=R(p_JG(x_J),q_JG(x_J)),\qquad
\f=S(p_JG(x_J),q_JG(x_J))\Eqa(A1.9)$$
where $R,S$ are defined above, see \equ(A1.6) which are written in a
form easily recognized in the elliptic functions tables.  The
functions $R,S$ can be rewritten in the following form:
\def\txt{\textstyle}
$$\eqalign{\txt
R(p,q)=&\txt-4g\left[\sum_{m=0}^\io\bigr({x^mp\over 1+x^{2m}p^2}
+{x^mq\over 1+x^{2m}q^2}\bigl)\right]\cr\txt
S(p,q)=&\txt4\left[\sum_{m=0}^\io\bigl(\atan x^mp-\atan
x^mq\bigr)\right]\cr
\txt
\sin {S(p,q)\over2}=&\txt
\fra{2g_J(x)}{g}                         
\left[\sum_{m=0}^\io (-1)^m\bigl(
{x^m p\over 1+ x^{2m} p^2}-{x^m q\over 1+x^{2m}q^2}\bigr)\right]\cr\txt
\cos {S(p,q)\over2}=&\txt
\fra{g_J(x)}{2g}                         
\left[1-2\sum_{m=0}^{\io}(-1)^m\bigl({x^{2m}p^2\over 1+x^{2m}p^2}+
{x^{2m}q^2\over 1+x^{2m}q^2}\bigr)\right]\cr}\Eqa(A1.10)$$
exhibiting some of the properties of the Jacobi map in a better way.

One checks that in the $(p_J,q_J)$ variables the pendulum hamiltonian,
\equ(A1.1) has become a function $J(p_Jq_J)=2g^2+ g x_J+O(x_J^2)$. The
domain of definition of the map is given by the properties of the
elliptic functions or, more restrictively, by the domain of
convergence of the above series. It inludes a disk of some radius
$\r_J>0$ around the origin.

The important symmetry $R(p,q)=R(q,p)$ and $S(p,q)=-S(q,p)$ is
manifest.
\*

\0{\bf Appendix A2. Filling times of quasi periodic motions.}
\numsec=2\numfor=1
\*

Let $(\o_1,\ldots,\o_d)=\oo\in R^d$ be such that
$|\oo\cdot\nn|^{-1}\le C |\nn|^\t$. Let $\ch(x), \ch_\perp(x)$ be
$C^\io$--functions even and strictly positive for $|x|<\fra12\p$,
vanishing elsewhere and with integral $1$. Let $\pps,\pps_0\in T^d$
and $x(\pps)\defi\e^{-(d-1)}\ch(\oo\cdot(\pps-\pps_0)/|\oo|)\cdot
\ch(\e^{-1}\,|P^\perp(\pps-\pps_0)|)$, $P^\perp=$ orthogonal
projection on the plane orthogonal to $\oo$.

The function $x$ can be naturally regarded as defined and periodic on
$T^d$: if $\hat\ch(\s)$ is the Fourier transform of $\ch$ as a
function on $R$ then the Fourier transform of $x$ is
$\hat\ch(\n^\parallel) \hat\ch(\e |\nn^\perp|)$, $\nn$ integer
components vector, $\n^\parallel=\oo\cdot\nn/|\oo|$,
$\nn^\perp=P^\perp\nn$.  The average $\lis
T^{\,-1}\ig_0^{\lis T} x(\oo t) \,dt$ is:

$$X=1+\sum_{\nn\ne\V0}\hat x(\nn) e^{-i\pps_0\cdot\nn}\fra1{\lis T}
\fra{e^{i\oo\cdot\nn \lis T} -1}{i \oo\cdot\nn}\ge
1-\fra{2C}{\lis T} \sum_{\nn\ne\V0}\Big|\hat\ch(\n^\parallel) \hat\ch(\e
|\nn^\perp|)\Big|\,|\nn|^\t\Eqa(A2.1)$$
Since the last sum is bounded above by $b \e^{-(\t+d-1)}$ the average $X$
is positive for all $\pps_0$ if $\lis T> 2bC\e^{-(\t+d-1)}$. This means
that for $T>2bC\e^{-(\t+d-1)}+\p/|\oo|$, hence for $T> B C
\e^{-(\t+d-1)}$ with $B$ a suitable constant depending only on $d$,
the torus will have been filled by the trajectory of any point within
a distance $\e$.  This proof is taken from (5), p. 111, of [G1], see
[BGL] for an alternative proof.
\*

{\bf References}
\*

\item{[A] } Arnold, V.: {\it Instability of dynamical systems with several
degrees of freedom}, Sov. Mathematical Dokl., 5, 581-585, 1966.

\item{[Be] } Bessi, U.: {it An approach to Arnold's diffusion through
the Calculus of Variations}, Nonlinear Analysis, 1995.

\item{[Br] } Bernard, P.: {\it Perturbation d'un hamiltonien
partiellement hyperbolique}, C.R. Academie des Sciences de Paris,
{\bf 323}, I, 189--194, 1996.

\item{[BGL] } Bourgain, J., Golze, ., Lochak, P.: {\it },
preprint, in print in J. Statistical Physics.

\item{[CG] } Chierchia, L., Gallavotti, G.: {\it Drift and diffusion in
phase space}, Annales de l' Institut Poincar\`e, B, {\bf 60}, 1--144,
1994.

\item{[E] } Eliasson, L.H.: {\it Absolutely convergent series expansions
for quasi-periodic motions}, Ma\-the\-ma\-ti\-cal Physics Elctronic Journal,
MPJE, {\bf 2}, 1996.

\item{[G1] } Gallavotti, G.: {The elements of mechanics}, Springer, 1983.

\item{[G2] } Gallavotti, G.: {\it Twistless KAM tori}, Communications in
Mathematical Physics {\bf 164}, 145--156, (1994).

\item{[Ge] } Gentile, G.: {\it A proof of existence of whiskered tori with
quasi flat homoclinic intersections in a class of almost integrable
systems}, Forum Mathematicum, {\bf 7}, 709--753, 1995. See also: {\it
Whiskered tori with prefixed frequencies and Lyapunov spectrum},
Dynamics and Stability of Systems, {\bf 10}, 269--308, 1995.

\item{[GG] } G. Gallavotti, G. Gentile: Majorant series convergence for
twistless KAM tori, {\sl Ergodic theory and dynamical systems} {\bf 15},
857--869, (1995).

\item{[GGM] } G. Gallavotti, G. Gentile, V. Mastropietro: {\it
Pendulum: separatrix splitting}, preprint, chao-dyn@xyz. lanl. gov
9709004. And G. Gallavotti, G. Gentile, V. Mastropietro: {\it Arnold's
diffusion is anisochronous systems}, in preparation.

\item{[GR] } Gradshteyn, I.S., Ryzhik, I.M.: {\sl Table of integrals series
and products}, Academic Press, 1965.

\item{[L] } Lochak, P.: {\it Arnold's diffusion: a compendium of
remarks and questions}, Proceedings of 3DHAM, s'Agaro, 1995, in print.

\item{[P] } Perfetti, P.: {\it Fixed point theorems in the Arnol'd model
about instability of the action--variables in phase space},
mp$\_$arc@math.utexas.edu, \#97-478, 1997, in print in
Discrete and continuous dynamical systems.

\item{[T] } Thirring, W.: {\it Course in Mathematical Physics}, vol. 1,
p. 133, Springer, Wien, 1983.

\*\*

\FINE
\*
\0Archived also in:\\
mp$\_$arc@math.utexas.edu        \# 97-???    and\\
chao-dyn@xyz.lanl.gov \kern0.4cm \# 9709011
\ciao